\documentclass[aps,prd,preprint,groupedaddress]{revtex4-1}

\usepackage{graphicx}

\begin{document}

\title{A New Class of Optical Beams for Large Baseline Interferometric Gravitational Wave Detectors}


\author{Stefan W. Ballmer}
\email[]{sballmer@syr.edu}
\affiliation{Syracuse University, Syracuse, New York, USA}

\author{David J. Ottaway}
\email[]{david.ottaway@adelaide.edu.au}
\affiliation{ Department of Physics and The Institute of Photonics and Advanced
Sensing, The University of Adelaide, Adelaide, South Australia, Australia}


\date{\today}

\begin{abstract}
A folded resonant Fabry-Perot cavity has the potential to significantly reduce the impact of coating thermal noise on the performance of kilometer scale gravitational wave detectors. When constructed using only spherical mirror surfaces it is possible to utilize the extremely robust $TEM_{00}$ mode optical mode. In this paper we investigate the potential thermal noise improvements that can be achieved for third generation gravitational wave detectors using realistic constraints. Comparing the previously proposed beam configurations such as e.g. higher order Laguerre-Gauss modes, we find that similar or better thermal noise improvement factors can be achieved, while avoiding degeneracy issues associated with those beams.
\end{abstract}


\maketitle

\section{Introduction}
\begin{figure}[!b]
\center
\includegraphics[width=3.2in]{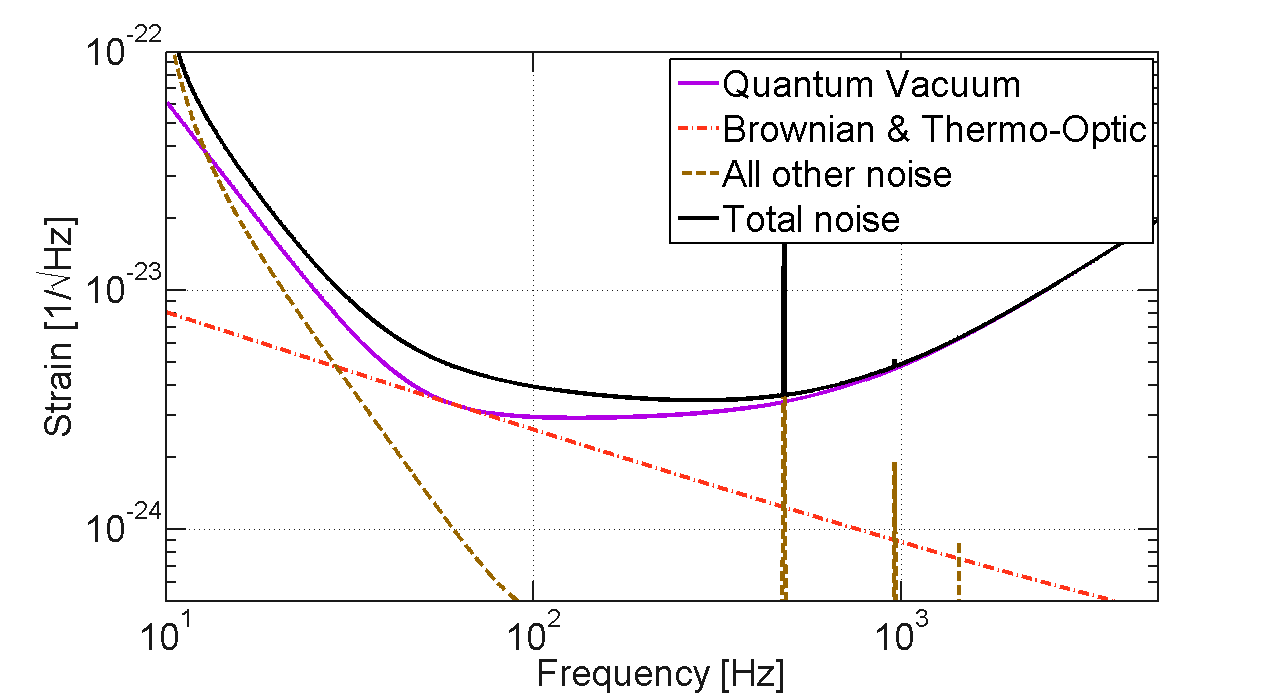}
\includegraphics[width=3.2in]{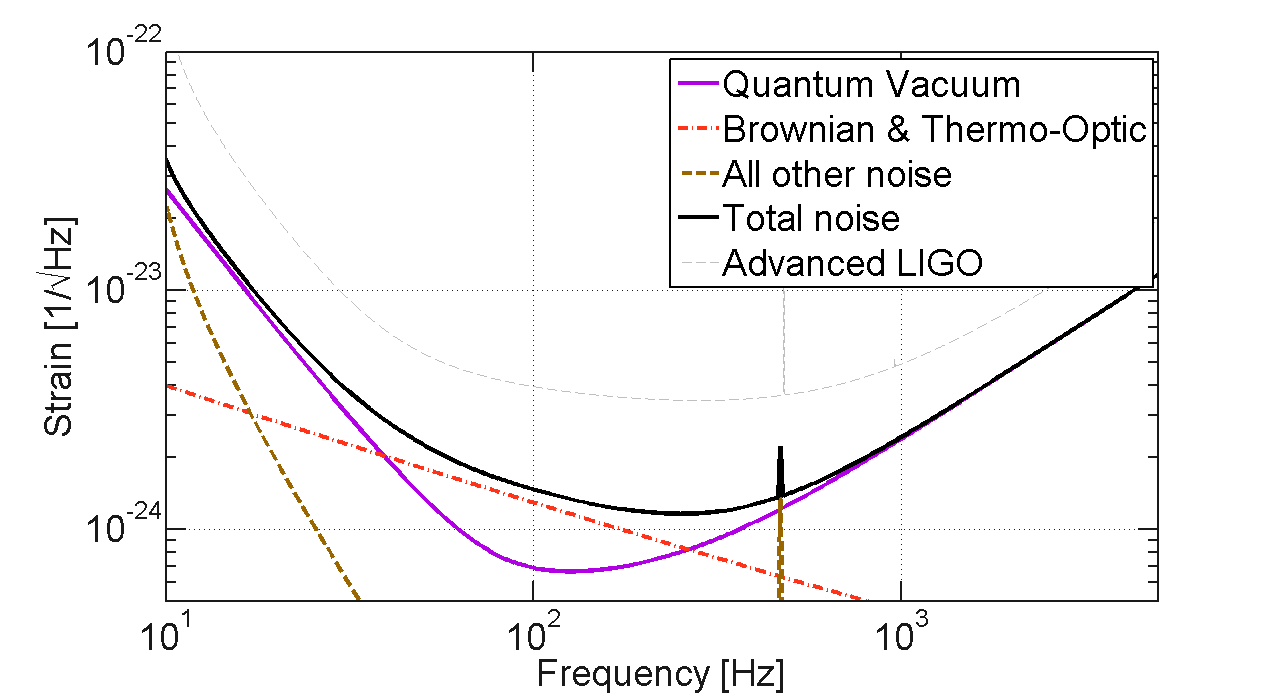}
\vspace*{-5pt}
\caption{Left: Simplified noise budget for Advanced LIGO, showing the relative contribution of quantum noise, mirror thermal noise (Brownian and Thermo-optic) as well as everything else to the design sensitivity. Right: Example noise budget for a modified LIGO interferometer with the 4.5 spot resonant delay line shown in figure \ref{fig:EllipSW}. The design uses existing Advanced LIGO mirror coatings. To improve the quantum noise the test masses were increased from $40~{\rm kg}$ to $160~{\rm kg}$, and a broadband quantum noise reduction of $6~{\rm dB}$ through the use of non-classical states of light was assumed. Finally all mirror reflectivities were then re-tuned to optimize the quantum noise in this configuration. Both plots were calculated using the Gravitational Wave Interferometer Noise Calculator (GWINC) tool, a software developed by the gravitational wave community.}
\label{fig:aL}
\vspace*{-10pt}
\end{figure}

The field of gravitational wave astrophysics is entering an exciting new phase. First generation long baseline interferometers have completed a year long integrated data run and a number of significant upper limits on astrophysical events have been placed \cite{abbott2009ligo, acernese2007status}. These first generation detectors are now being replaced by second generation instruments which are anticipated to have sensitivity that is improved by almost and order of magnitude and a low frequency cut-off that is reduced from 40 Hz to 10 Hz or below \cite{harry2010advanced, virgo2007,kuroda2010status}. It is anticipated that the sensitivity of these instruments will be limited by a combination of thermal noise and quantum noise which is due to the quantum nature of light used to read the interferometer out, see figure \ref{fig:aL}. Research is now being conducted into possible configurations that are suitable for third generation instruments that will be begin construction around 2020 \cite{punturo2010einstein}.

To make significant enhancement on the sensitivity of second generation instruments two stubborn noise sources must be addressed, namely quantum noise and coating thermal noise. The use of non-classical states of light and filter cavities looks like a promising method to
reduce the quantum noise
of second generation interferometers by a factor of at least three \cite{kimble2001conversion,collaboration2011instrument}. The maximum benefit of this noise improvement will only be achieved if a corresponding improvement is realized in coating thermal noise.

Coating thermal noise is the name given to the noise caused by the mechanical dissipation of the dielectric coatings applied to the test masses to create high reflectivity surfaces. The impact of coating thermal noise on advanced gravitational wave detectors was first realized over 10 years ago \cite{PhysRevD.57.659,harry2002thermal,crooks2002excess}. Since that time significant amount of research has been directed into determining the cause of the mechanical dissipation in coatings and finding new coatings to reduce it\cite{martin12}. Despite nearly a decade of work, these heroic efforts have improved the coating thermal noise of amorphous coatings by $31$ percent\cite{martin12}. Recently crystalline coatings have been demonstrated that reduce coating thermal noise by a factor of over three in amplitude \cite{cole2013tenfold}. This impressive demonstration was performed using a coating whose spatial extent was significantly less than 25mm. Considerable effort is still needed to determine whether this coating technology can be scaled to the size needed for third generation optics whilst maintaining this thermal noise improvement and all of the other demanding specification required of such a coating.

In parallel with developing new coatings other researchers have been looking at new interferometer topologies to reduce the impact of coating thermal noise. Nakagawa et al. showed that delay lines instead of Fabry-Perot arms can significantly reduce the impact of thermal noise because of the improved averaging of different spots reflecting from mirrors \cite{nakagawa2002estimating}. Other researchers have investigated using higher order Laguerre Gaussian modes \cite{mours2006thermal,PhysRevD.79.122002} and non-spherical mirrors such as the so called "sombrero" \cite{bondarescu2006new} and conical mirrors \cite{bondarescu2008optimal}. $TEM_{00}$ Gaussian modes have been shown to have a low sensitivity to mirror perturbations compared to these other mirror geometries and mode types, see for example \cite{PhysRevD.84.102001,miller2008thermal}. We have therefore sought out a mirror geometry that can maintain the advantages offered by resonating $TEM_{00}$ modes.

Optical delay lines were first proposed by Herriott \cite{Herriott1964}.
They were incorporated in the original paper on interferometric gravitational wave detection by Weiss \cite{WeissQPR72}. In this paper we present an investigation of resonant delay lines, also called folded Fabry-Perot cavities, for use in long baseline gravitational wave detectors. Our approach uses a modified Herriott Delay line approach to fold a Fabry-Perot cavity many times to increases the sampling of the mirror surface and hence reduce coating thermal noise. The technique however also has possible applications for low noise reference cavities.   The use of resonant delay lines for improving the sensitivity of the optical readout of resonant mass detectors has been investigated by Marin et al.\cite{marin2003folded,anderlini2009kg} In this paper we address the issues that are relevant to applying this technique to multi-kilometer interferometric gravitational wave detectors. The techniques discussed here can be applied in isolation or in combination with improvements to the coatings or with other beam techniques such as the implementation of Laguerre gauss modes.

\section{Coating thermal noise}

Second generation gravitational-wave detectors are limited by thermal noise in their most sensitive band. Specifically the limiting source of noise is the Brownian motion of the mirror surface, caused by the mechanical loss in the mirror coating. According to \cite{PhysRevD.57.659} the power spectral density of this noise is given by
\begin{equation}
S_{xx} = \frac{4 k_B T}{ \pi f} \phi U
\label{TN1}
\end{equation}
where $\phi$ the loss angle, and U the strain energy associated with a static pressure profile on the mirror surface, normalized by the total driving force. Specifically for Brownian noise due to mechanical loss in the coatings, read out by a Gaussian $\rm TEM_{00}$ mode, $U$ is given by \cite{Harry:06,lrr-2009-5}
\begin{equation}
U = \delta_c \frac{(1+\sigma)(1-2\sigma)}{\pi Y w^2} \Omega_1
\label{TN2}
\end{equation}
where $\delta_c$ is the coating thickness, $\sigma$ is the Poisson ratio of the substrate, $Y$ is the Young's modulus of the substrate and $w$ is the Gaussian beam width, i.e. the laser intensity is proportional to $\exp(-2 r^2/w^2)$. Finally $\Omega_1$ is a correction factor with $\Omega_1=1$ if coating and substrate have the same Young's modulus and Poisson ratio. Corrections for higher order optical modes and finite size mirrors have also been calculated \cite{lrr-2009-5}. Equations \ref{TN1} and \ref{TN2} show that the coating thermal noise is the Brownian surface motion of the optic, averaged over the laser beam spot area.

In addition we are interested in the spatial correlation of thermal noise across the mirror surface. For the dominant coating Brownian noise the intrinsic spatial correlation drops off over about the coating thickness $\delta_c$, which for almost all applications is much small than the  Gaussian beam width $w$. Note that since the coating Brownian noise calculation depends on the non-local elastic Greens function, the statement above is not necessarily obvious, and will break down at the first internal resonance frequency. The interested reader is referred to a paper by G. Lovelace \cite{lovelace2007dependence}.
The noise correlation of neighboring beam spots with width $w$ at locations $x_A$ and $x_B$ therefore scales with the optical overlap
\begin{equation}
S_{x_A x_B} = S_{x x} \frac{\int I(\vec{r}-\vec{x}_A) I(\vec{r}-\vec{x}_B) d^2r}{\int I^2(\vec{r}) d^2r} = S_{x x} e^{-\frac{|\vec{x}_A-\vec{x}_B|^2}{w^2}}
\label{eq:correlation1}
\end{equation}
Here we used the intensity profile of the beam $I(r)\propto \exp(-2r^2/w^2)$.
In addition, recently proposed crystalline coating materials may be limited by thermo-optic noise  \cite{cole2013tenfold}. Here the spatial correlation is dictated by the frequency dependent diffusion length
\begin{equation}
d_{\rm diff}= \sqrt{\frac{\kappa}{2 \pi f C \rho}}
\end{equation}
with $\kappa$ the thermal conductivity, $C$ the specific heat and $\rho$ the density.
As long as $d_{\rm diff}$ is much smaller than the Gaussian beam width $w$, equation \ref{eq:correlation1} is equally valid for thermo-optic noise. Note that this condition might be violated for certain cryogenic reference cavities. Equation \ref{eq:correlation1} is however not valid
for Brownian noise due to mechanical loss in the mirror substrate. For this case
Nakagawa et. al. \cite{nakagawa2002estimating} showed that the spatial correlation  is dictated by the elastic Greens function of the optic, and is given by
\begin{equation}
S^{\rm SB}_{x_A x_B} = S^{\rm SB}_{x x} e^{-\frac{|\vec{x}_A-\vec{x}_B|^2}{2w^2}} I_0 \left( \frac{|\vec{x}_A-\vec{x}_B|^2}{2w^2} \right)
\label{eq:correlation2}
\end{equation}
where $I_0$ is the modified Bessel function of the first kind.
Since substrate Brownian noise is significantly below the coating noise however, we do not have to be concerned with this correlation.

The Advanced LIGO coatings are silica-tantala ($\rm Si O_2$-$\rm Ta_2 O_5$) dielectric stacks with a titania-doping  ($\rm Ti O_2$) in the $\rm Ta_2 O_5$ layers. The same coatings are also intended to be used in the Advanced Virgo detector. They were selected for low mechanical loss, while respecting the additional optical specifications. The dominant mechanical loss is due to the high-index $\rm Ta_2 O_5$. The titania-doping of the $\rm Ta_2 O_5$ layers is the main improvement over the initial LIGO coatings. It resulted in a reduction of the loss angle from about $4 \times 10^{-4}$ \cite{Harry:06} in undoped coatings to about $2.5 \times 10^{-4}$ with the titania
doping \cite{Harry:2006qh}.
Ideas to further improve the coating noise roughly fall into three classes:
(i) Selecting an alternative coating material with optimized material constants. One can directly target the mechanical loss angle. This approach was chosen in \cite{cole2013tenfold} by using crystalline coatings. The titania doped tantulum coatings for Advanced LIGO also fall into that category. Alternatively one could  optimize other parameters. For instance one could aim for a larger refractive index contrast, allowing for thinner, lower noise coatings.
(ii) Switch to cryogenic operation to reduce thermal noise. While equation \ref{TN1} suggest linear improvement of the noise power spectral density with temperature, the temperature dependence of other material parameters - in particular the mechanical loss angle - can negate any benefit. This is for example the case for the substrate material of choice at room temperature - fused silica ($\rm Si O_2$). Thus, to benefit from cryogenic operation of an interferometer, a change of substrate and coating material is also required.
(iii) Effectively sample a larger mirror area. Resonant delay lines offer to do the latter while staying away from extremely degenerate optical cavities. The rest of the paper will explore this approach.

\section{Resonant Delay Lines}

\begin{figure}[!b]
\center
\includegraphics[width=4.5in]{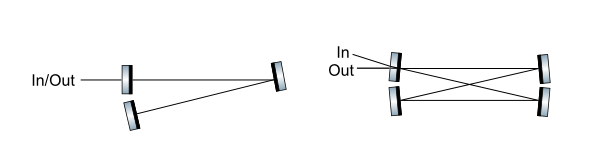}
\vspace*{-5pt}
\caption{Fabry-Perot cavities configured in standing wave (left) and traveling wave geometries (right)}
\label{fig:SW_Travelling_Wave}
\vspace*{-10pt}
\end{figure}

A resonant delay line can be operated in either a traveling wave or a standing wave configuration. Figure \ref{fig:SW_Travelling_Wave} illustrates simple Fabry-Perot cavities configured in both standing wave and traveling wave configurations. The traveling wave geometry clearly has advantage of separating the inputs and outputs. This may be advantageous for introducing squeezed vacuum into the interferometers because the squeezed vacuum need not be introduced through an optical isolator with losses that degraded the potential benefit. This benefit comes with considerable cost in the form of significantly increased complexity of the system.

The choice of standing wave or traveling wave also effects the thermal noise improvement that can be achieved for a given arm length. In Table \ref{table:twsw} we calculate the relationship between the number of bounces on each mirror, the resulting increase in the round trip cavity length and improvement in coating thermal noise for the simple case in which all spot sizes are identical and they are sufficiently separated that their thermal noise contribution is uncorrelated to that from adjacent spots. For a traveling wave interferometer this means that the thermal noise contribution of each spot is equal. However in a standing wave geometry the cavity mode samples the intermediate spots on the mirror twice, picking up the same thermal noise twice coherently. This leads to a slight more complicated scaling of thermal noise with increasing bounce number.

\begin{table}
\begin{center}
\begin{tabular}{| l | c | c | }
\hline
Parameter & Traveling-wave & Standing-wave \\ \hline
Spots per mirror  & $N_b$ & $N_b$ \\
Total reflections & $2 N_b$ & $4 N_b -2$ \\
Round trip length & $2 N_b L$ & $(4 N_b -2) L$ \\
Displacement amplitude thermal noise factor & $\sqrt{2 N_b}$ & $\sqrt{8 N_b -6}$ \\
Strain amplitude thermal noise factor & $\frac{1}{\sqrt{2 N_b}}$ & $\frac{\sqrt{8 N_b -6}}{4 N_b -2} = \frac{1}{\sqrt{2 (N_b -\frac{1}{4}) + \frac{1}{8 N_b -6}}}$ \\
Thermal noise reduction factor & $ \sqrt{N_b}$ & $\frac{4 N_b-2}{\sqrt{16 N_b-12}} =
\sqrt{N_b-\frac{1}{4} + \frac{1}{16 N_b - 12}}
$ \\
\hline
\end{tabular}
\end{center}
\caption{Comparison between traveling-wave and standing-wave configurations. Listed are total number of reflections, total round trip length, and amplitude thermal noise scaling factors (defined as thermal noise ratio between one mirror reflection and the complete configuration). For simplicity we assume here that all spots are the same size, and neighboring spots are completely uncorrelated. All expressions are given in terms of the number of spots per mirror, $N_b$, and the cavity length $L$. For the standing-wave configuration $N_b$ can be half-integer, indicating one additional beam spot on the input coupler. The thermal noise reduction factor is the improvement in thermal noise compared to a standard Fabry-Perot cavity with identical spot sizes.}
\label{table:twsw}
\end{table}

Resonant delay lines increase the sensitivity of the interferometer because the thermal noise adds incoherently, whereas the gravitational wave signal add coherently for each additional pass. This coherent addition of the gravitational wave signal is equivalent to making the arms of the interferometer longer, which increases the sensitivity to gravitational wave strain at lower frequencies. However increasing the round trip length of the arm cavities also reduces their free spectral range. A gravitational wave detector with a folded Fabry-Perot arm cavity is insensitive to gravitational-wave strain at its free spectral range, as we will see below.

Hence we are not able to increase the arm length arbitrarily. Further, higher order cavity modes may create undesired resonances in the detector and these also reduce in frequency as the arm cavity length increases and the free spectral range decreases.

For an interferometer with 4km long beam tubes like Advanced LIGO this argument limits the effective increase in arm cavity round-trip  length to approximately 10 times which would put the free spectral range and its resulting zero in interferometer response at 3.75 kHz. This sets the number of bounces to 4 bounces on the ETM and 5 on the ITM for a standing wave and 10 on each for a traveling wave cavity. This means a theoretical improvement of 2.1 and 3.2 in the strain sensitivity for standing wave and traveling wave cavities respectively compared to a conventional Fabry-Perot cavity with the same spot size at each bounce. From this point of view traveling wave cavities have a significant advantage. However this comes at the cost of a significantly more complicated layout of power and signal recycling cavities in a gravitational-wave interferometer. A traveling wave interferometer does have the small advantages of lessening the requirements on optical isolation and provides a convenient port for injecting non-classical states of light for squeezed light enhancement of gravitational wave detectors.

\section{A Simple Ring Cavity Solution}
\label{sec:sphericalRDL}
\begin{figure}[!b]
\center
\includegraphics[width=2.8in]{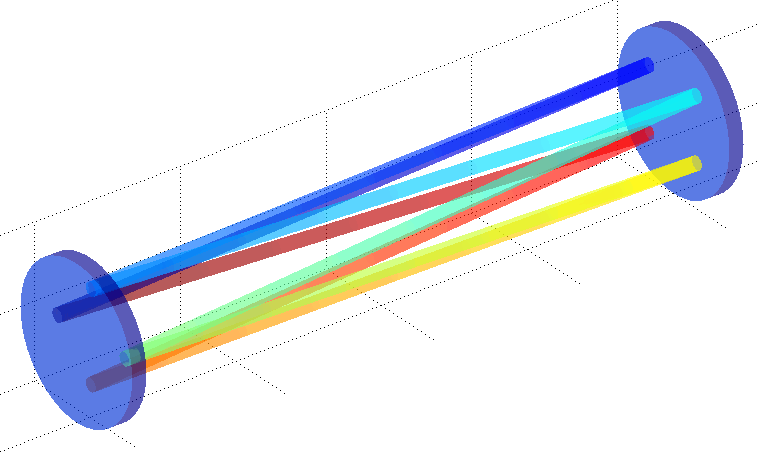}
\includegraphics[width=2.8in]{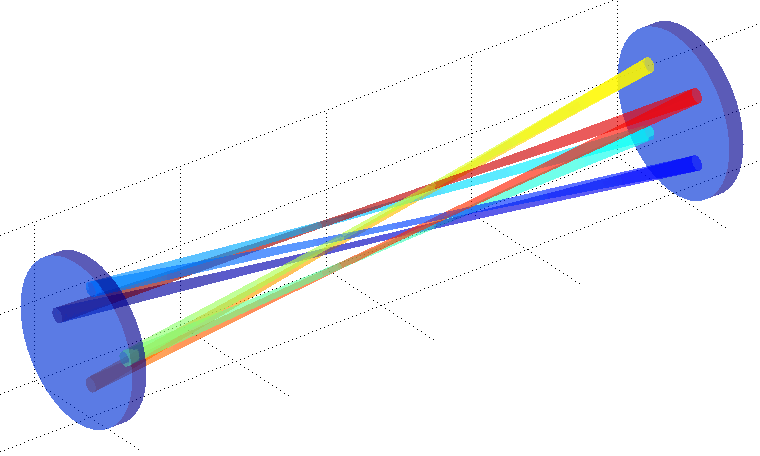}
\vspace*{-5pt}
\caption{Example of a positive branch (left) and negative branch (right) folded Fabry-Perot cavity with 4 bounces. Both represent a traveling-wave configuration.}
\label{fig:posnegbranch}
\vspace*{-10pt}
\end{figure}

Our goal is to reduce thermal noise by effectively averaging over more of the mirror surface, without introducing the instabilities associated with using a single large spot. Further, by using mirrors with a spherical curvature wherever possible the task of polishing the optics becomes easier. The first geometry that we consider here is a simple pair of spherical mirrors illustrated in Figure \ref{fig:posnegbranch}. The solutions that we looked for was a Herriott delay line path in which the final bounce connected with with the input beam position and upon reflection from the mirror became aligned with the input beam and hence closed the path forming a resonant delay line.
We use geometrical optics and ray transfer matrices to trace the folded optical axis.

The round trip ray transfer matrix M of a Fabry-Perot cavity is given by:
\begin{equation}
M = \left(
\begin{array}{ccc}
A & B \\
C & D
\end{array}
\right)=
\left(
\begin{array}{ccc}
1 & 0 \\
-\frac{2}{R_{itm}} & 1
\end{array}
\right)
\left(
\begin{array}{ccc}
1 & L \\
0 & 1
\end{array}
\right)
\left(
\begin{array}{ccc}
1 & 0 \\
-\frac{2}{R_{etm}} & 1
\end{array}
\right)
\left(
\begin{array}{ccc}
1 & L \\
0 & 1
\end{array}
\right)
\label{equ:M_matrix}
\end{equation}
where $R_{itm}$ and $R_{etm}$ denote the radius of curvature of the input test mass (input mirror) and end test mass (far mirror) respectively. The product of its eigenvalues is $\lambda_1 \lambda_2 = 1$, since
the determinant of M is $1$.
The optical stability criterion for this cavity is $|\lambda_i| \leq 1$ for all eigenvalues, which implies $\lambda_1 =e^{i \phi}$ and $\lambda_2=\lambda_1^*$. \cite{siegman1986lasers}. This is a sufficient condition for an optical mode to be present, independent of the choice of $\phi$.
A resonant delay line mode however needs to repeat itself exactly after $N_{b}$ bounces per mirror, leading to additional constraints on the angle $\phi$. In particular, after $N_{b}$ bounces per mirror we need to fulfill the condition
\begin{equation}
M^{N_{b}}
\left(
\begin{array}{ccc}
x \\
x'
\end{array}
\right)
= 1
\left(
\begin{array}{ccc}
x \\
x'
\end{array}
\right)
\label{equ:M_idempotent}
\end{equation}
with $x$ the beam position and $x'$ the beam slope. This implies $M^{N_{b}}$ has one eigenvalue of $1$, and since the determinant is also $1$, all eigenvalues are 1. A beam path that connects correctly with its self therefore requires $M^{N_{b}}$ to be the identity matrix, and equation \ref{equ:M_idempotent} is true for any input beam $(x;x')$. For the eigenvalues of $M$ this implies
\begin{equation}
\lambda_1 =e^{i \frac{2\pi n}{N_b}}, \,\,\,\,\,\,\,\, n=(1,2,...,N_b)
\label{equ:eigenvaldef}
\end{equation}
and $\lambda_2=\lambda_1^*$. Using the cavity g-factors $g_{itm}=1-L/R_{itm}$ and $g_{etm} = 1-L/R_{etm}$ this can be expressed as (for details see appendix \ref{app:gg})
\begin{equation}
g_{itm}g_{etm} = \frac{\Re{(\lambda_1)}+1}{2}=\cos^2(\frac{\pi n}{N_b}),  \,\,\,\,\,\,\,\, n=(1,2,...,N_b)
\label{equ:cond}
\end{equation}
Equation \ref{equ:cond} is identical to the condition for the $N_b$-th order transverse modes to be co-resonant in the cavity with the fundamental mode, highlighting the connection between a higher-order transverse mode and a folded beam path in a two mirror cavity with spherical mirrors. It implies that $M^{N_{b}}$ is the identity matrix, which presents a problem for two reasons. First, our folded cavity is completely degenerate, and has a $0~{\rm Hz}$ transverse mode spacing. In other words it has no mode selection ability, similar to short plane-parallel cavity (etalon). Second, equation \ref{equ:cond} is a marginally stable point design. Any slight deviations in either radius of curvature or cavity length will result in a cumulative drift of consecutive reflections, destroying the mode shape.

We therefore conclude that the simple spherical mirror design has to be modified. We are interested in a minimal modification, preserving simple spherical mirrors for most of the beam spots, for two reasons: (i) having the same spherical shape for neighboring spots will reduce the clipping loss on reflection, and (ii) retaining the overall spherical shape of the mirrors will reduce the complexity of manufacture.

Locally modifying the radius of curvature for a single reflection (e.g. on the input test mass) does not lead to a stable cavity configuration. This can be seen by calculating the modified round trip ray transfer matrix $M_{r.t.}^{mod}$, and remembering that $M^{N_{b}}$ is the identity matrix:
\begin{equation}
M_{r.t.}^{mod}=
\left(
\begin{array}{ccc}
1 & 0 \\
-\frac{2}{R_{itm}^{mod}} & 1
\end{array}
\right)
\left(
\begin{array}{ccc}
1 & 0 \\
-\frac{2}{R_{itm}} & 1
\end{array}
\right)^{-1}
M^{N_{b}}
 =
\left(
\begin{array}{ccc}
1 & 0 \\
-\frac{2}{R_{itm}^{mod}} + \frac{2}{R_{itm}}  & 1
\end{array}
\right)
\label{equ:M_matrix_mod}
\end{equation}
which has geometric multiplicity of 1 and is not optically stable. This constraint does not hold if we modify the radius of curvature for two reflections, which can be shown by an example. Therefore a stable, folded, traveling-wave optical cavity can be achieved by locally polishing shallow cups or by perturbing two locations on the spherical mirrors using thermal compensation \cite{lawrence2003active}, thus perturbing the otherwise spherical mirrors for a total of two spots per cavity.

To maximize the gain of a folded cavity design the spot size of each reflection must be kept as big as possible. The ray transfer matrix analysis above shows that there is an inherent connection between spot size and number of bounces $N_b$ if the path is to close on itself. The spot size on the mirrors in a traditional Fabry-Perot two-mirror cavity is given by
\begin{equation}
w_1^2 = \frac{\lambda L}{\pi} \sqrt{\frac{g_2}{g_1(1-g_1 g_2)}}
\label{equ:spotsize1}
\end{equation}
and $w_2^2=w_1^2  g_1/g_2$. The individual spot sizes for a folded cavity will vary slightly around that number due to the radius of curvature perturbation that need to be introduce at two locations to ensure mode discrimination. For the symmetric case $g_1=g_2$, and using equation \ref{equ:cond} we find
\begin{equation}
w_{1,2}^2 = \frac{\lambda L}{\pi} \sqrt{\frac{1}{(1- \cos^2(\frac{\pi n}{N_b}))}}   \,\,\,\,\,\,\,\, n=(1,2,...,N_b)
\label{equ:spotsizesym}
\end{equation}
Maximizing the beam waist $w_{1,2}$ therefore leads us to pick $n=1$ (or equivalently $n=N_b-1$, since $\lambda_2=\lambda_1^*$, see equation \ref{equ:eigenvaldef}). $n=1$ implies that we are stepping through neighboring spot on a mirror, not skipping any spots. We still have the choice of picking either a positive or a negative g-factor, see figure \ref{fig:posnegbranch}. Both lead to the same spot sizes and thermal noise, but negative g-factor configurations have been preferred in the 2nd generation of gravitational wave detectors since they lead to lower angular optical spring frequencies \cite{sidles2006optical}.

\begin{figure}[!b]
\center
\includegraphics[width=4in]{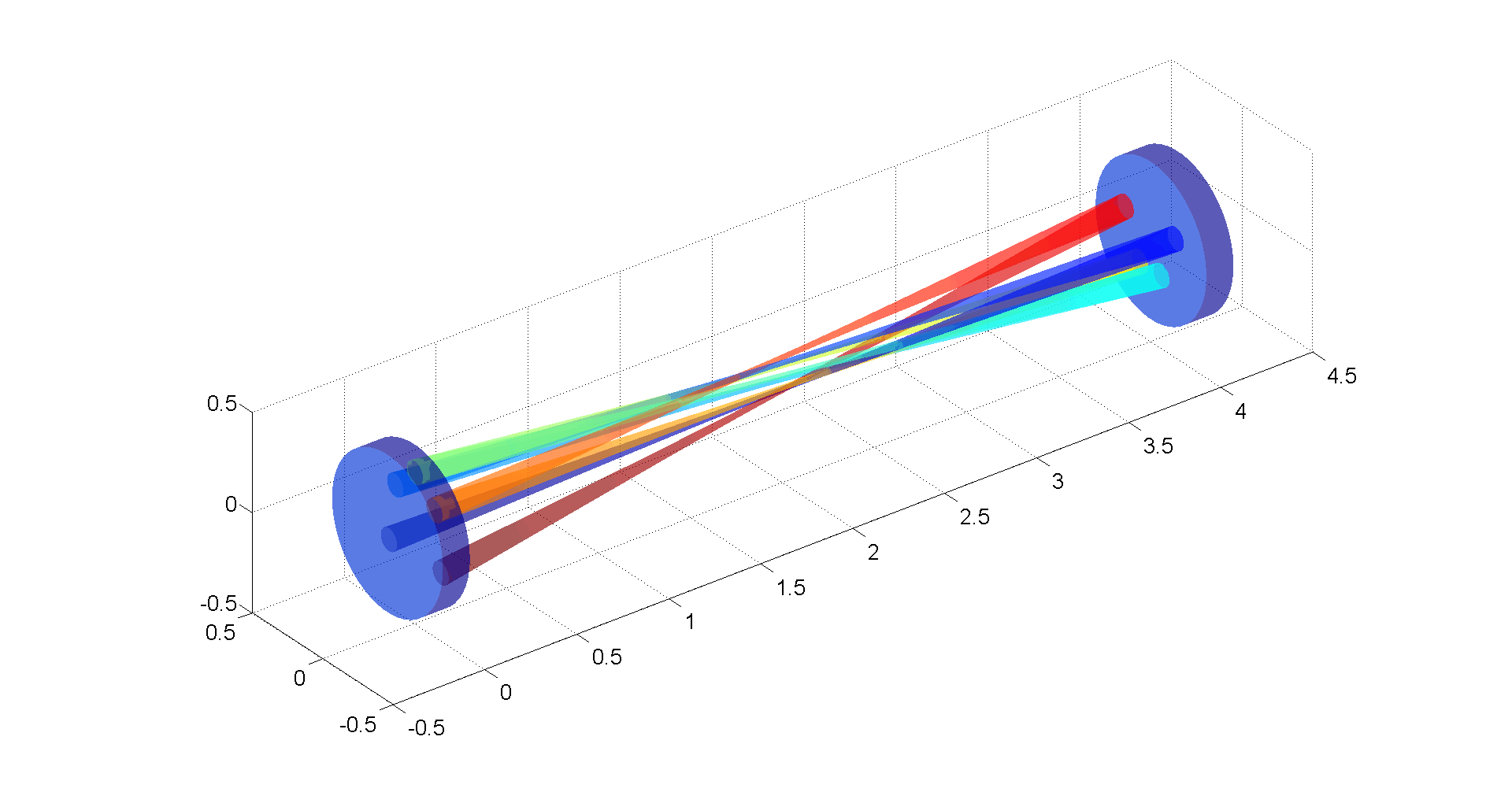}
\includegraphics[width=2.3in]{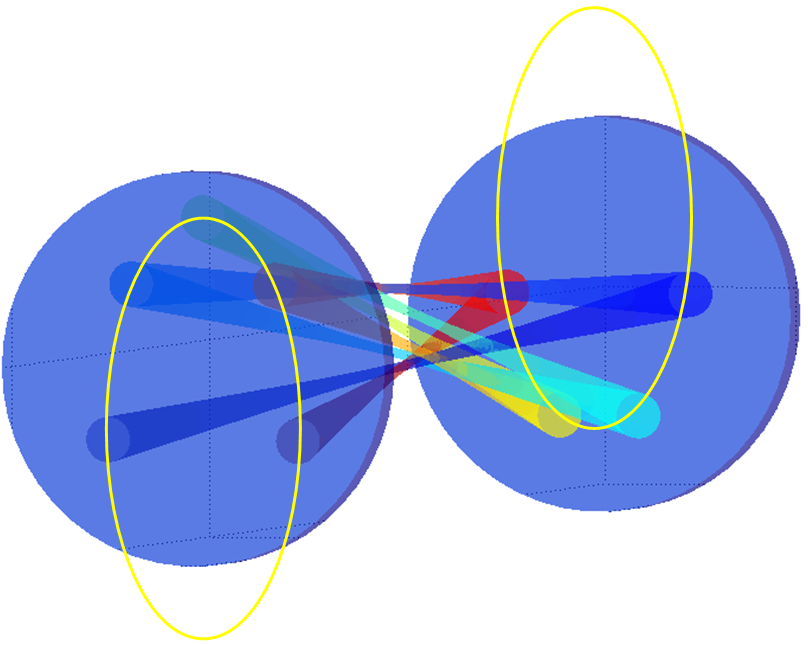}
\vspace*{-5pt}
\caption{Standing wave cavity with an elliptical beam orbit, 5 spots on the input test mass, and 4 spots on the end test mass. Left: side view; right: frontal view with the elliptical beam orbits indicated in yellow. The g-factor was chosen to be equal to Advanced LIGO ($g=0.8303$), corresponding to $N_b=7.4$ bonces per orbit, and a beam size of $w=57.3~{\rm mm}$. The optics have a radius of $40~{\rm cm}$, and a thickness of $15~{\rm cm}$, corresponding to a mass of $160~{\rm kg}$ if made out of fused silica. This design provides a beam clearance of about $2.5 w$, limiting clipping losses to about $1~{\rm ppm}$ (lower on the end test mass). The coating Brownian thermal noise of this configuration is $2.1$ times below Advanced LIGO in amplitude. (The arm length is shrunk by a factor 1000 for illustration purpose.)}
\label{fig:EllipSW}
\vspace*{-10pt}
\end{figure}

The design spot sizes for Advanced LIGO are $w=53~{\rm mm}$ and $w=62~{\rm mm}$ for the input and end test mass respectively, corresponding to a design cavity g-factor of $g_{itm}g_{etm}=0.8303$ (negative branch). This corresponds to an effective bounce number $N_b=7.4$ per mirror (equation \ref{equ:cond}). For a traveling wave geometry this is achievable with a mirror of about 1m which is approaching the maximum clear aperture of the LIGO beam tubes. The spacing in this case is driven by clipping loss requirements on the input and output coupling surfaces which is discussed later in this paper.

Standing wave geometries require that a significant wedge is built on two areas on the optic where the terminating bounces of the cavity hit. This means that the design is no longer tied into meeting the stringent requirements described above. Thus the g-factor is not fixed. Further it allows the design to consider possible configurations with less than one full orbit. One interesting approach is to choose an elliptically shaped beam orbits (e.g. the yaw beam oscillations smaller than the pitch oscillations), and terminate the orbit with wedges after roughly 1/2 orbit.
This configuration has multiple advantages:
\begin{itemize}
 \item We can choose a g-factor that results in comparable or bigger beam spots than current gravitational wave interferometers.
\item We can pick the number of actual spots to match the available mirror surface area.
\item By placing the wedged input and output coupler near the minor half-axis of the beam orbit, we can maximize the spot separation for those beams, thus minimizing clipping losses at the wedged surfaces.
\item All other spots will be closer to each other, but they are all supported by the same mirror radius of curvature, thus no clipping loss will occur.
\end{itemize}
Figure \ref{fig:EllipSW} shows such an elliptical standing wave cavity with 4.5 beam spots (4 on the end test mass, five on the input test mass). Its g-factor is identical to Advanced LIGO's. It guarantees at least $2.5 w$ of beam clearance from mirror edges or wedged area edges, keeping clipping losses down. An example interferometer sensitivity that can be achieved with this design is shown in figure \ref{fig:aL}.

\section{Alignment Control}
The behaviour of the described resonant delay line under misalignment of a mirror is surprisingly simple. The Gaussian beam propagating down the cavity has the same $q$-parameter as the mode of a simple two-mirror Fabry-Perot cavity. Here the q-parameter is defined as $1/q = 1/R -i \lambda/( \pi w^2)$ where R is the wavefront radius of curvature, and w is the spot size.  Therefore the spot motion under misalignment is identical to the simple Fabry-Perot cavity case, and is given by
\begin{equation}
\left(
\begin{array}{c}
x_1\\
x_2
\end{array}
\right)
=
\frac{L}{1-g_1 g_2}
\left(
\begin{array}{cc}
-g_2 & 1\\
-1 & g_1
\end{array}
\right)
\left(
\begin{array}{c}
\theta_1\\
\theta_2
\end{array}
\right)
\end{equation}
As a consequence traditional Pound-Drever-Hall wavefront sensing can be used for cavity or input beam alignment control, just as in the simple Fabry-Perot cavity case \cite{fritschel1998alignment,barsotti2010alignment}.

\section{Spot Spacing Limitations}

\begin{figure}[!b]
\center
\includegraphics[width=4.5in]{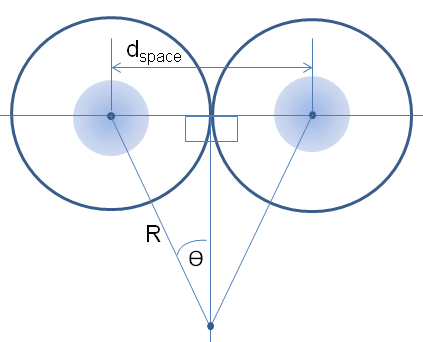}
\vspace*{-5pt}
\caption{The geometry used to calculate the mirror size required to support an $N_{b}$ bounces per mirror delay line}
\label{fig:Mirror_Size}
\vspace*{-10pt}
\end{figure}

We discussed in the previous section how coating thermal noise is limited if the maximum area of the mirror is interrogated by light. Given that the correlation length of coating thermal noise is very short this generally does not set a significant constraint on the spacing of the spots. However the combination of the need to locally alter the mirrors on at least two locations and the need to effectively outcouple the beam from the cavities does. In aLIGO the mirror diameter is set by the requirement to the limit diffraction losses per bounce. This means that the aLIGO mirrors have a radius that is 2.7 times the spot radius of the fundamental beam radius. To maintain the same loss per bounce means the spot separation must be twice this amount if the spots are evenly spaced. The elliptical pattern described earlier relaxes this constraint somewhat because the spot spacing near the terminal bounces is larger than the intermediate bounces.

For the ring delay line described earlier it is simple to determine an analytical relationship between the spot size, number of bounces, desired separation and the mirror size. The geometry used to describe this situation is illustrated as Figure \ref{fig:Mirror_Size}. In this figure $d_{space}$ is the desired separation between the spots. Half this distance will be the clear aperture around the spots. The angle $\Theta$ is given in radians by $\Theta = \frac{\pi}{N_{b}}$. Using trigonometry we find $d_{space}/2 = R \sin \Theta$, and the required mirror diameter is simply:
\begin{equation}
D_{mirror} = 2(R+d_{space}/2) = d_{space}(1+\frac{1}{sin(\frac{\pi}{N_{b}})})
\label{equ:Mirror_Size}
\end{equation}

In practice this equation is a more severe limitation for the traveling wave geometry than the FSR restriction described earlier. The clear aperture of the LIGO vacuum system is 1 m which limits the number of bounces to 6 if the same effective clear mirror area to beam size as LIGO is maintained. This will enable a coating thermal noise of 2.45 in amplitude to be achieved. The additional freedom that a standing wave geometry allow means that a more compact spacing of spots can be achieved by allowing the input/output bounces to be spread further apart than the bounces near the center. The other complication that results from significantly increasing the diameter of the test mass is that the solid-body modes of the test mass reduce in frequency. It will be necessary to consider this as part of any future detector design because this can have limit the high frequency performance of the detector.

 \section{Scattered Light Control}
Early prototypes of gravitational wave detectors utilized multi-bounce delay lines as an alternative of to Fabry-Perot cavities. The low frequency performance limits of these detectors was often attributed to scattered light \cite{shoemaker1988noise}. Since these early experiments there has been considerable improvement on the achievable mirror surface quality and a dramatic enhancement in seismic isolation available. However it is prudent to do some analysis to determine whether scattered light is likely to set significant additional requirements on the control of the optics in a resonant delay line cavity. It is also worth pointing out that in a resonant delay line the net round-trip length of the cavity will be fixed which is not necessarily the case for a Michelson or Sagnac interferometer with conventional delay lines. Further given the stable mirror geometries that are used it should be possible to resonate an auxiliary laser between the center of the two mirrors in a single bounce standing wave Fabry-Perot geometry similar to the green laser that will be used in aLIGO \cite{brooks2009auxiliary,mullavey2011arm}. This will allow the accurate control of the microscopic separation and angular alignment of these cavities. During the next part of this section we will evaluate the requirements on angular stability of these mirrors to prevent scattered light from becoming a performance limitation.

For scattered light to be an issue, light must first be scattered from one site, be incident on the location of another bounce and then be scattered back into the mode exiting the mirror at the new location. The fraction of light that makes this transition was evaluated by Flanagan and Thorne \cite{flanagan1994noise}. and for this case is given by:

\begin{equation}
\delta I/I = (\frac{\lambda}{L})^{2} BRDF_{mirror}(\theta_{exit})BRDF_{mirror}(\theta_{recomb})
\label{equ:Scattered_Light}
\end{equation}

Light that is re-injected in this manner does not necessarily reduce the sensitivity of the instrument unless it picks up additional time dependent phase shifts. One way that this can occur is if there is angular motion of the test masses. In this situation the effective displacement noise that this creates is:

\begin{equation}
S_x^{1/2} = \sqrt{\delta I/I}\Delta x S_\theta^{1/2}
\label{equ:Scattered_Light_Recomb}
\end{equation}

where $\Delta x$ is the difference in spacing between where the mode hits the mirror correctly and the spot in which the scattered light recombines and $S_\theta^{1/2}$ is the angular amplitude displacement spectra. It is constructive to compare this equation with the coupling of angular noise to a standard Fabry-Perot cavity in which the locations of where the beams bounce of the mirrors are offset from the center of the mirror by an amount $\Delta x_{disp}$ is given by the formula:

\begin{equation}
S_x^{1/2} = \Delta x_{disp} S_\theta^{1/2}
\label{equ:Angular_sensitivity}
\end{equation}

It is expected that the mis-centering tolerance for aLIGO will be $50\mu m$ compared with the maximum spot separation which in a resonant delay line could be 0.5 m. However this noise term from scattering is considerably attenuated by the coupling between the two paths. Using the polished aLIGO mirrors as a guide. The BRDF can be as high as 3000, which when plugged into Equation \ref{equ:Scattered_Light} gives a value of $5 \times 10^{-13}$ which makes this term smaller than the conventional angular noise by a factor of 100.

It is well known that if re-injected scattered light takes a path whose length is modulated by greater than the wavelength of light then sidebands are imposed on the light can have a frequency separation from the carrier that is considerably larger than the frequency of the original path modulation. This is so-called unconverted noise (see \cite{ottaway2012impact} for example). To ensure that this is not an issue the maximum RMS angular fluctuations of the mirror must be limited to less $\lambda/d_{mirror}$ = $1\mu Rad$ which is considerably more than anticipated for aLIGO when under active control.


\section{Gravitational wave antenna function}
\label{FSRsensitivity}
Folding a gravitational wave interferometer arm also affects its transfer function for gravitational waves. For light traveling down an arm (aligned with the x-axis) and back, the change in round trip time delay is given by
\begin{equation}
\delta T_1(\omega) = D(\omega, n_x) h_{xx}(\omega)
\label{equ:rtt1}
\end{equation}
where $h_{xx}(\omega)$ is the stain component along the x-arm (in transverse-trace-less gauge), $n_x$ the component along the arm of the normal vector pointing at the souce (i.e. opposite to the gravitational wave k-vector). The transfer function $D(\omega, n_x) $ is given by \cite{Malik}
\begin{equation}
D(\omega, n_x) = \frac{1}{-2 i \omega} \left[ \frac{1-e^{i \omega (1-n_x)T}}{1-n_x}
- e^{2 i \omega T} \frac{1-e^{-i \omega (1+n_x)T}}{1+n_x} \right]
\label{equ:rtttf}
\end{equation}
Here $T$ is the one-way light travel time in the arm ($4~{\rm km}/c$ for Advanced LIGO).
We now fold the beam $N_{r.t.}$ times, with $N_{r.t.}=N_b$ for traveling-wave geometry and $N_{r.t.}=2 N_b - 1$ for standing-wave geometry. Equation \ref{equ:rtt1}  becomes
\begin{equation}
\delta T_{N_{r.t.}}(\omega) = D(\omega, n_x) F_{N_{r.t.}}(\omega) h_{xx}(\omega)
\label{equ:rtt2}
\end{equation}
with
\begin{equation}
F_{N_{r.t.}}(\omega) = 1 + e^{i 2 \omega T} + ... + e^{i 2(N_{r.t.}-1) \omega T} = \frac{e^{i 2 N_{r.t.} \omega T}-1}{e^{i 2 \omega T}-1}
\label{equ:rttF}
\end{equation}
As expected we have $F_1(\omega)=1$ and $F_{N_{r.t.}}(\omega) \rightarrow N_{r.t.}$ for $\omega \rightarrow 0$.
However, in contrast to a regular Fabry-Perot cavity, the sensitivity to gravitational waves of a folded cavity at its free spectral range $\omega_{\rm FSR}= {\pi}/{(N_{r.t.}T)}$ is exactly zero. Since a gravitational wave interferometer should have good sensitivity up to a few kHz, this constrains the total number of reflections $N_{r.t.}$ to less than  about 10 for an arm length of $4~{\rm km}$ (Advanced LIGO).

\section{Conclusion}
We have presented an analysis of a new topology for future gravitational wave detectors that reduces the impact of coating thermal noise by a factor of up to 2.5 in amplitude. This new topology improves the averaging of coating thermal noise across the surface of the test masses. The proposed design makes use of lowest order Gaussian beams which have been shown to be the most stable optical mode against imperfections in mirror surfaces. The topology can also be used in conjunction with improvements in mirror coatings such as the recently developed crystalline coatings, and thus has the potential to elimitate coating thermal noise as principal design constraint for gravitational wave interferometer sensitivity.
The challenges for implementing this topology include the need for relatively large test masses, as well as the unusual polishing requirement to achieve a different radius of curvature at certain spots.

\section{Acknowledgement}
This work was supported by NSF grant PHY-1068809 and the Australian Research Council(ARC).
The LIGO document control center number for this paper is LIGO-P1300082.

\section{Appendix: g-factor constraint for resonant delay line}
\label{app:gg}
Here we revisit the two-mirror resonant delay line with spherical mirrors first discussed in section \ref{sec:sphericalRDL}. Our starting point is equation \ref{equ:M_matrix}, describing the two-mirror Fabry-Perot cavity round trip ray transfer matrix $M$. In the main text we have seen that for a resonant delay line with $N_b$ bounces per mirror, the eigenvalues $\lambda_i$ of $M$ have to fulfill equation \ref{equ:eigenvaldef}. Using the cavity g-factors $g_{itm}=1-L/R_{itm}$ and $g_{etm} = 1-L/R_{etm}$, we can express $M$ as
\begin{equation}
M =
\left(
\begin{array}{ccc}
1 & 0 \\
\frac{2}{L}(g_{itm}-1) & 1
\end{array}
\right)
\left(
\begin{array}{ccc}
1 & L \\
0 & 1
\end{array}
\right)
\left(
\begin{array}{ccc}
1 & 0 \\
\frac{2}{L}(g_{etm}-1) & 1
\end{array}
\right)
\left(
\begin{array}{ccc}
1 & L \\
0 & 1
\end{array}
\right)
\label{equ:M_matrixApp}
\end{equation}
or
\begin{equation}
M =
\left(
\begin{array}{ccc}
2 g_{etm} -1 & ... \\
 ... & 2 g_{itm} -2 + (2 g_{itm} -1)(2 g_{etm} -1)
\end{array}
\right)
\label{equ:M_matrixApp2}
\end{equation}
Using $\lambda_2=\lambda_1^*$ (eq. \ref{equ:eigenvaldef}) we can relate the real part of $\lambda_1$ to the trace of $M$:
\begin{equation}
\Re{(\lambda_1)} = \frac{1}{2} (\lambda_1 + \lambda_2) =  \frac{1}{2} tr(M) = 2 g_{itm}g_{etm} -1
\label{equ:TrM}
\end{equation}
Finally, since $\Re{(\lambda_1)} = \cos(\frac{2 \pi n}{N_b})$ (eq. \ref{equ:eigenvaldef}), and using the trigonometric identity
\begin{equation}
\cos^2(\alpha) = \frac{\cos(2 \alpha) + 1}{2}
\label{equ:TrigId}
\end{equation}
we find the condition for a two-mirror resonant delay line with spherical mirrors (eq. \ref{equ:cond})
\begin{equation}
g_{itm}g_{etm} = \frac{\Re{(\lambda_1)}+1}{2}=\cos^2(\frac{\pi n}{N_b}),  \,\,\,\,\,\,\,\, n=(1,2,...,N_b)
\label{equ:condRepeat}
\end{equation}

\bibliography{Ottaway_Ballmer_V6}

\end{document}